\documentclass{article}
\usepackage{arxiv}
\usepackage{multirow}
\usepackage{amssymb}
\usepackage{amsmath}
\usepackage{float}
\usepackage[utf8]{inputenc} % allow utf-8 input
\usepackage[T1]{fontenc}    % use 8-bit T1 fonts
\usepackage{hyperref}       % hyperlinks
\usepackage{url}            % simple URL typesetting
\usepackage{booktabs}       % professional-quality tables
\usepackage{amsfonts}       % blackboard math symbols
\usepackage{nicefrac}       % compact symbols for 1/2, etc.
\usepackage{microtype}      % microtypography
\usepackage{lipsum}		% Can be removed after putting your text content
\usepackage{graphicx}
\usepackage{natbib}
\usepackage{doi}

\title{Slice-by-slice deep learning aided oropharyngeal cancer segmentation with adaptive thresholding for spatial uncertainty on FDG PET and CT images}

\author{
\href{https://orcid.org/0000-0003-0455-9795}{\includegraphics[scale=0.06]{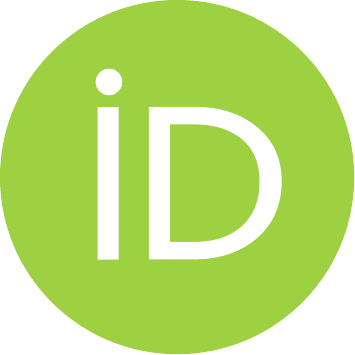}\hspace{1mm}Alessia De Biase}\thanks{\texttt{a.de.biase@umcg.nl}} \\
	Department of Radiation Oncology\\
	University Medical Center Groningen\\
	Groningen, NL 9700RB \\
	\And
\href{https://orcid.org/0000-0001-6644-274X}{\includegraphics[scale=0.06]{orcid.pdf}\hspace{1mm}Nanna Maria Sijtsema} \\
	Department of Radiation Oncology\\
	University Medical Center Groningen\\
	Groningen, NL 9700RB \\
	\And
\href{https://orcid.org/0000-0002-9515-5616}{\includegraphics[scale=0.06]{orcid.pdf}\hspace{1mm}Lisanne van Dijk} \\
	Department of Radiation Oncology\\
	University Medical Center Groningen\\
	Groningen, NL 9700RB \\
	Department of Radiation Oncology\\
	University of Texas MD Anderson Cancer Center\\
	Houston, TX 77030 \\
   \And
\href{https://orcid.org/0000-0003-1083-372X}{\includegraphics[scale=0.06]{orcid.pdf}\hspace{1mm}Johannes A. Langendijk} \\
	Department of Radiation Oncology\\
	University Medical Center Groningen\\
	Groningen, NL 9700RB \\
	\And
\href{https://orcid.org/0000-0002-8995-1210}{\includegraphics[scale=0.06]{orcid.pdf}\hspace{1mm}Peter van Ooijen} \\
	Department of Radiation Oncology\\
	Data Science Center in Health (DASH)\\
	University Medical Center Groningen\\
	Groningen, NL 9700RB \\
}

\date{}

\renewcommand{\headeright}{}
\renewcommand{\undertitle}{}
\renewcommand{\shorttitle}{}

\hypersetup{
pdftitle={Slice-by-slice deep learning aided oropharyngeal cancer segmentation with adaptive thresholding for spatial uncertainty on FDG PET and CT images},
pdfsubject={q-bio.NC, q-bio.QM},
pdfauthor={Alessia De Biase},
pdfkeywords={Automatic tumor segmentation, Adaptive Segmentation, Deep Learning},
}

\begin{document}
\maketitle

\begin{abstract}
Tumor segmentation is a fundamental step for radiotherapy treatment planning. To define an accurate segmentation of the primary tumor (GTVp) of oropharyngeal cancer patients (OPC), simultaneous assessment of different image modalities is needed, and each image volume is explored slice-by-slice from different orientations. Moreover, the manual fixed boundary of segmentation neglects the spatial uncertainty known to occur in tumor delineation. This study proposes a novel automatic deep learning (DL) model to assist radiation oncologists in a slice-by-slice adaptive GTVp segmentation on registered FDG PET/CT images.
We included 138 OPC patients treated with (chemo)radiation in our institute. Our DL framework exploits both inter and intra-slice context. Sequences of 3 consecutive 2D slices of concatenated FDG PET/CT images and GTVp contours were used as input. A 3-fold cross validation was performed three times, training on sequences extracted from the Axial (A), Sagittal (S), and Coronal (C) plane of 113 patients. Since consecutive sequences in a volume contain overlapping slices, each slice resulted in three outcome predictions that were averaged. In the A, S, and C planes, the output shows areas with different probabilities of predicting the tumor. The performance of the models was assessed on 25 patients at different probability thresholds using the mean Dice Score Coefficient (DSC). Predictions were the closest to the ground truth at a probability threshold of 0.9 (DSC of 0.70 in the A, 0.77 in the S, and 0.80 in the C plane). The promising results of the proposed DL model show that the probability maps on registered FDG PET/CT images could guide radiation oncologists in a slice-by-slice adaptive GTVp segmentation. 
\end{abstract}

% keywords can be removed
%Automatic tumor segmentation, Adaptive Segmentation, Deep Learning

\section{Introduction}
Image-guided deep learning tools have shown enormous potential in the field of radiation oncology. Two of the main goals are to speed up the radiotherapy workflow and to automate error prone tasks. Thus, automatic segmentation of organs at risk (OARs) and clinical target volumes (CTVs) is of great interest, especially in cases where various imaging modalities are needed for radiotherapy treatment planning. For the head and neck region, 3D deep learning segmentation models on Positron Emission Tomography (PET) and Computed Tomography (CT) images were designed and tested during the first and second edition of the HECKTOR challenge \cite{Andrearczyk2020d}\cite{Andrearczyk2022}. The achieved results, in terms of Dice Similarity Coefficient (DSC) and $95^{th}$ Hausdorff Distance, showed the potential of Artificial Intelligence (AI) in the tumor segmentation task.

In recent years, analysis of the variability of target volume definitions across different and same observers, documented as inter- and intra-observer variability, have been published \cite{Sadeghi2021}\cite{Weiss2003}. Several studies have investigated the extent of inter-observer variability in target volume delineation, showing large discrepancies in the contoured structures \cite{interobserver2019}. 
Automatic tumor segmentation has demonstrated to achieve lower performance compared to automatic organs at risk delineation, since tumor volumes highly depend on the clinical scenario and clinical judgement of the treating physician \cite{Comparing2020}. During the learning process of automatic segmentation models, manual annotations of the volume of interest are used as ground truth. As contouring varies from specialist to specialist, the use of manual contours as a reference may not be reliable \cite{Sadeghi2021}, thus these variations can cause lack of generalization of the method.

Different strategies to improve tumor delineation and reduce inter-observer variability have been proposed. Some examples are consensus guidelines, reliable ground truth, the use of proper and complementary imaging modalities \cite{Gudi2017}\cite{Sadeghi2021} or discuss the contours in a multidisciplinary meeting. The use of FDG PET/CT for target volume delineation, for example, showed to significantly reduce inter-observer variability for head and neck cancers compared to using CT only \cite{Gudi2017}. In addition, the absence of a proper reference representing the absolute gold standard makes it difficult to establish which contour can be chosen for a certain clinical task and to quantify such variability. Despite the above mentioned solutions may improve tumor delineation, they are not always the most efficient. Adding more image modalities, in fact, is expected to improve the contouring accuracy 
\cite{RodriguezOuteiral2021}, but makes the contouring process more time consuming and more complex \cite{Anderson2014}.

In this paper we propose a 2D deep learning segmentation method that shows the degree of uncertainty of each pixel to be classified as tumor or not on CT and PET images of oropharyngeal cancer patients. The deep learning framework takes sequences of three consecutive slices from the CT and PET images as input. The corresponding sequences extracted from the gross tumor volume (GTV) manually delineated by experts are used as ground truth. Each slice in a volume can be either first, second, or third in a sequence, resulting in three different predictions of the same slice. In the axial, coronal, and sagittal view, the model outputs probability maps instead of binary masks. Bi-directional long short term memory (Bi-LSTM) and spatial and channel attention mechanisms are used to capture context information from the nearby slices and to enhance inter-slice dependencies. 
With our method, we would like to assist radiation oncologists in decision making, giving them a deep learning aided tool for tumor segmentation in radiotherapy treatment planning, where the optimal threshold for clinical acceptance could be customized and where the uncertainty derived from the variability in contouring in the training set is available and displayed.

\section{Materials and methods}

\subsection{Data collection}

Data used in this study were manually collected from the Picture Archiving and Communication System (PACS) of the University Medical Center of Groningen (UMCG) in the Netherlands retrospectively. We considered oropharyngeal cancer patients (OPC) who were treated, in our institute, with (chemo)radiation therapy between 2014 and 2017. 166 patients satisfied the eligibility criteria.
For each case, planning CT and FDG PET 3D images, and GTV primary tumor delineations, that were used in the radiotherapy treatment planning process, were collected. All imaging was acquired in treatment position using fixation with a mask. The tumor delineations were reviewed by a specialized head and neck nuclear medicine physician and by a head and neck radiologist in cases where the MRI scan was also available. The tumor contours were approved after a multidisciplinary second reading session attended by all head and neck radiation oncologists. 

Firstly, rigid image registration was performed. Radiation oncologists' manual annotations of primary tumors were extracted from the DICOM RTSTRUCT files. Then, the imaging data were downloaded in DICOM format and transformed into NIfTI file format \cite{M.D.1995} to facilitate data handling. PET images were transformed from raw to Standardized Uptake Values (SUV) \cite{Ulaner2019}. For 9 patients, the PET image could not be registered with the planning CT, because either the PET image was missing or the images co-registration was not accurate due to differences in patient posture between the PET and CT images. Only patients having one primary tumor and who received radiation therapy as the first cancer treatment were included in this study. 
The final group satisfying the inclusion criteria consisted of 138 OPC patients (age: 61.95 (mean) ± 9.02 (std) years). Participants provided written informed consent for their data to be used for research purposes. In Table \ref{distribution}, the dataset description including gender, T and N stages, and HPV status is reported.

\begin{table}
\centering
\caption{Dataset description (n=138)}\label{distribution}
\centering
\vspace*{0.3cm}
\begin{tabular}{lrrr}
\hline
%\vspace*{0.5cm}
%\rule{0pt}{2ex}   
\multicolumn{2}{c}{\multirow{3}{*}{\bfseries Variables}} & Training & Test \\
\multicolumn{2}{c}{} & (n=113) & (n=25) \\
\multicolumn{2}{c}{} & \emph{n (\%)} & \emph{n (\%)}  \\

\hline

\rule{0pt}{3ex}    
Gender   & male      & \small{69} \footnotesize{(61)} &  \small{18} \footnotesize{(72)}  \\
         & female     & \small{44} \footnotesize{(39)} &   \small{7} \footnotesize{(28)}  \\

T Stage  & T1      & \small{16} \footnotesize{(14)} &  \small{5} \footnotesize{(20)} \\
         & T2     & \small{30} \footnotesize{(27)} &   \small{4}  \footnotesize{(16)} \\
         & T3    & \small{13} \footnotesize{(11)} &   \small{3} \footnotesize{(12)} \\
         & T4    & \small{54} \footnotesize{(48)} &  \small{13} \footnotesize{(52)}  \\
             
N Stage  & N0      & \small{18} \footnotesize{(16)} &   \small{5} \footnotesize{(20)} \\
         & N1     & \small{13} \footnotesize{(11)} &   \small{0}  \footnotesize{(0)} \\
         & N2a    & \small{6} \footnotesize{(5)} &   \small{0} \footnotesize{(0)} \\
         & N2b    & \small{37} \footnotesize{(33)} &  \small{8} \footnotesize{(32)}  \\
         & N2c    & \small{36} \footnotesize{(32)} &  \small{10} \footnotesize{(40)}  \\
         & N3    & \small{3} \footnotesize{(3)} &  \small{2} \footnotesize{(8)}  \\

HPV status & negative    & \small{47} \footnotesize{(42)} &  \small{13} \footnotesize{(52)}  \\
         & positive    & \small{47} \footnotesize{(42)} & \small{9}  \footnotesize{(36)}  \\
         & unknown    & \small{19} \footnotesize{(16)} &  \small{3} \footnotesize{(12)}  \\
 
\hline
\end{tabular}
\end{table}

\subsection{Region of Interest selection and pre-processing}

%add specifics about PET and CT image dimensions and pixel dimensions
Medical Imaging data can be considerably large in dimensions. Since this study focused specifically on oropharyngeal cancer, a region of interest centered in the oral cavity was automatically extracted for each patient. The method used for the bounding box selection is inspired by \cite{Andrearczyk2020f}. It consists of a two step approach: first, brain segmentation is performed identifying the brain as the largest connected component in the image after applying a fixed threshold for the SUV values of three in the PET image; second, a bounding box of fixed size (144x144x144) is determined with respect to the brain position, as in \cite{Andrearczyk2020f}. We observed that using a fixed threshold of three to separate the brain from the rest of the head did not work for some of the patients in our dataset. The main problem occurred in cases where tumor activity on the PET image was included as it was part of the brain. To avoid manual assessment of the quality of the bounding boxes, we identified some metrics to guide us in manually correcting the threshold for the SUV values. We considered the brain volume (expected to be around 1260 $cm^3$ in men and 1130 $cm^3$ in women) and the maximum, the mean and the standard deviation of the Hounsfield units (HU) in the corresponding CT image of the segmented area of the brain. High values of maximum HU suggest that the segmentation is too large because bones have a much larger HU compared to soft tissues. 
Once the correct bounding boxes had been identified, CT and PET images were reconstructed into 144x144x144 pixels resulting in volumes of 144x144x144 $mm^3$. 

To facilitate tumor segmentation, a level of 40 HU and a window of 350 HU were chosen for the CT images. The PET images pixel values below 0 were set to 0. Finally, for both PET and CT images z-normalization was performed.

%level=40
%window=350
%PET zscore norm and clip 0

\subsection{Proposed Deep Learning framework}

The proposed method is divided into two main sections: in the first part, a deep learning network was trained on PET and CT images to identify tumor areas based on image features; in the second part, the uncertainty of the predictions of the first step was reconstructed for each patient using probability maps.

\begin{figure}

\hspace*{-0.2cm}\includegraphics[scale=0.62]{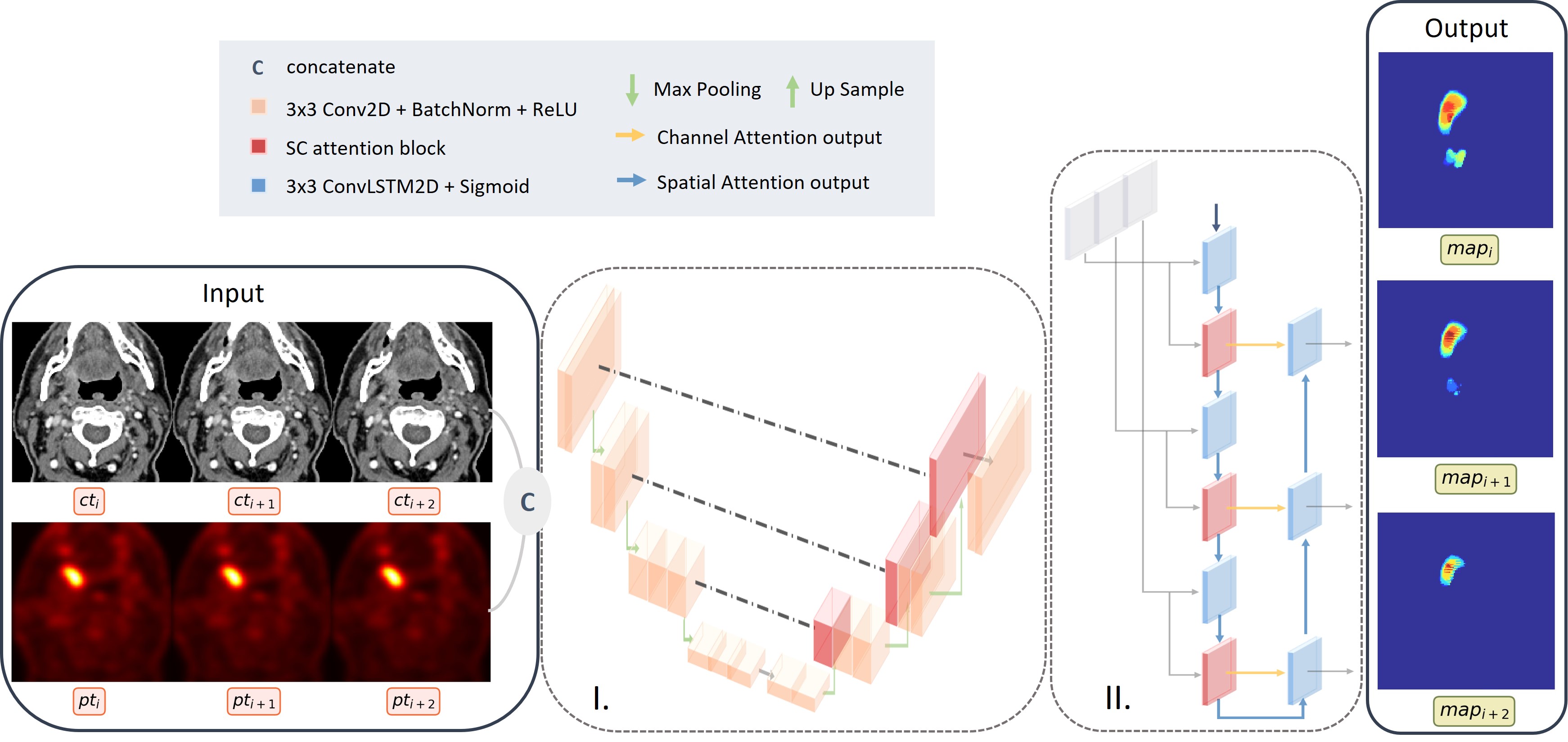}

\caption{The proposed deep learning framework. On the left side of the image the input data is displayed as two concatenated sequences of three slices from the PET [$pt_i, pt_{i+1}, pt_{i+2}$] and from the CT [$ct_i, ct_{i+1}, ct_{i+2}$] 3D volumes of a patient in the training set. The method is divided into two sections: {\bfseries I.} An encoder decoder path, comprising spatial and channel attention blocks; {\bfseries II.}  A bi-directional LSTM with spatial and channel attention blocks. The output of the framework is a reconstructed sequence of the three slices showing a probability map indicating the pixel tumor probability.} \label{workflow}

\end{figure}

\subsubsection{Prediction model on 2D sequences}

The proposed deep learning segmentation model is inspired by \cite{Pan2021c}. As training data, we used sequences of three consecutive 2D slices of PET [$pt_i, pt_{i+1}, pt_{i+2}$] and CT images [$ct_i, ct_{i+1}, ct_{i+2}$], concatenated in the channel domain (as shown in figure \ref{workflow}), together with three corresponding consecutive 2D slices of GTV primary tumor segmentation [$gtv_i, gtv_{i+1}, gtv_{i+2}$], in the form of binary masks. The main reason to use only three slices was computational costs.

Tumor segmentation, as many medical imaging problems, suffers of class imbalance. In a 2D model, the problem is intensified, since negative slices represent the majority of the volume. To train and validate the network, sequence selection was needed. A sequence [$i,  i+1, i+2$] was always selected if its first slice $i$ contained tumor pixels for more than $5\%$ of the total area. Only $10\%$ of the total number of sequences were selected if its first slice was either a negative slice ($5\%$) or it contained tumor pixels for less than $5\%$ of the total area (the other $5\%$). During the testing phase, all sequences from a volume were selected, containing both positive and negative slices. In order to recognize negative examples as such, we also needed to include them in the training set.

The framework is divided into two main steps: first, an encoder decoder path, comprising spatial and channel attention blocks, extracts the main image features from each sequence; second, a bi-directional LSTM with spatial and channel attention aims to capture context information from the nearby slices. The network is described in all its parts in figure \ref{workflow}. 

Loss functions were used as in \cite{Pan2021c}. Dice Score Coefficient (DSC) was calculated on the validation set to monitor the training process. To perform this task, a threshold of 0.5 was applied to the pixel probability values of the predictions to transform segmentation results into binary masks. We saved only three different checkpoints: one corresponding to the last trained model after 150 epochs, the second one corresponding to the model with the highest mean DSC on the validation set, and the final one corresponding to the model which obtained the second highest value of mean DSC after the first 40 epochs, indicating the point where fluctuations in the loss functions decreased.

\subsubsection{Single slice probability prediction}

As shown in figure \ref{workflow}, the output of the proposed network is a sequence [$map_i, map_{i+1}, map_{i+2}$] obtained as result of a sigmoid function, meaning that each pixel of the image corresponds to a probability indicating the chance of it belonging to tumor or not. During the testing phase, from each volume, a total of 142 sequences of three consecutive slices were extracted and predictions were calculated. For each slice in a volume, a minimum of one (for slice 1 and 144) and a maximum of three (from slice 3 to 142) different predictions were obtained. This is because slice number $k$, with $k$ from 3 to 142, will be first, second or third in a sequence. 
\hspace*{-1.5cm}
\begin{equation} \label{eq:1}
\begin{aligned}
    seq_{1} = \Big[map_{1}, map_{2}, map_{3}\Big] \quad\\
    \dots \quad\\
    seq_{k-1} = \Big[map_{k-2}, map_{k-1}, map_{k}\Big] \quad\\
    seq_{k} = \Big[map_{k-1}, map_{k}, map_{k+1}\Big] \quad\\
    seq_{k+1} = \Big[map_k, map_{k+1}, map_{k+2}\Big] \quad\\
    \dots \quad\\
    seq_{142} = \Big[map_{142}, map_{143}, map_{144}\Big] \quad\\\end{aligned}
\end{equation}

Lastly, each final predicted slice of a volume $out_k$ was calculated as the average of the total number of predictions for that slice: 

\begin{equation} \label{eq:2}
out_k = \begin{cases}
seq_{k}[1], & \scriptsize{k = 1}\\[8pt]
\frac{1}{2} (seq_k[1] + seq_{k-1}[2]), & \scriptsize{k = 2} \\[8pt]
\frac{1}{3}\sum_{i=1}^{3} seq_{k+1-i}[i], & \scriptsize{3 \leqslant k \leqslant 142} \\[8pt]
\frac{1}{2} (seq_{k-1}[2] + seq_{k-2}[3]), &  \scriptsize{k = 143} \\[8pt]
seq_{k-2}[3], & \scriptsize{k = 144} \\[8pt]
\end{cases}
\end{equation}

\hspace*{1cm}

The idea behind this last step is showing the uncertainty of the network in classifying each pixel as tumor to the end user. During testing on the hold-out test set, model ensembling was performed, averaging predictions obtained from the three models trained and validated during 3-fold cross validation. Ensemble learning combines the predictions from multiple deep learning models to reduce the variance of predictions and reduce generalization error \cite{Logan2021}. 

%out_k = seq_{k-1}[3]+seq_{k}[2]+seq_{k+1}[1] \\[10pt]

%\subsubsection{Input Output data}

%\subsubsection{Deep learning framework}

\subsection{Evaluation metrics and techniques}

The first step of the evaluation procedure was based on the quality of the predicted sequences. In order to evaluate the method, binary images were needed. A threshold was chosen for the pixel probability values of the predictions. For the second step of the evaluation procedure, we reconstructed each single slice from the predicted sequences of probabilities. For each patient we obtained 144 slices of probabilities for the axial, 144 for the sagittal and 144 for the coronal view. We calculated the metrics slice by slice and then we averaged over each patient at different probability thresholds. 

To evaluate the quality of the segmentation model, the mean DSC was used. Dice Score coefficient is largely used in image segmentation tasks \cite{Gudi2017}, because it quantifies the overlap between the prediction result and the ground truth image, which in our case is the GTV of the primary tumor. The DSC was calculated with the following formula on 2D images:

\begin{equation} \label{eq:3}
DSC = \frac{2 \times TP + \varepsilon_{smooth}}{(TP + FP) + (TP + FN) + \varepsilon_{smooth}}
\end{equation}

where TP indicated the number of true positives, FP the number of false positives and FN the number of false negatives.
A smoothing factor of $\varepsilon_{smooth} = 1 \times 10^{-5}$ was used to avoid errors in cases of well predicted sequences or slices not containing any tumor pixel.

Both precision and recall were used during the testing phase to point out the precarious outcomes. Precision defines how correct is the model when predicting that a specific area contains the tumor, while recall gives a percentage of the amount of tumor that has been identified by the model for a certain patient. Both these metrics are used to have an overall idea of the quality of the predictions, but both of them are not completely reliable if considered separately. After each final predicted slice $out_k$ of a volume was calculated, precision and recall were obtained with the following formulas:

\begin{equation} \label{eq:4}
Precision = \frac{\sum_{i=1}^{144} TP_{out_{i}}}{\sum_{i=1}^{144} TP_{out_{i}} + \sum_{i=1}^{144} FP_{out_{i}}}
\end{equation}

\begin{equation} \label{eq:5}
Recall = \frac{\sum_{i=1}^{144} TP_{out_{i}}}{\sum_{i=1}^{144} TP_{out_{i}} + \sum_{i=1}^{144} FN_{out_{i}}}
\end{equation}

%Both these metrics are used to have an overall idea of the quality of the model, but both of them are not completely reliable if considered separately. 
%Hausdorff Distance is complementary to DSC since it assesses the quality of the segmentation outline in terms of distance. The lowest the HD is, the better.

%\subsubsection{Deep learning model evaluation}

\section{Experimental results}

\subsection{Implementation details}

%CORRECT NUMBER OF PATIENTS FROM THE METHOD PART

The deep learning framework was implemented and trained using Python (v3.7.4), PyTorch (v1.6.0), MONAI (v0.8.0), CUDA (v10.1), and cuDNN (v7.6.4) on GPUs of the Peregrine Cluster of the University of Groningen. The training and validation set consisted of 113 patients, while the test set consisted of 25 patients. The same network was trained in parallel three times using, each time, different input data: sequences from the axial ($M_a$), sagittal ($M_s$) and coronal ($M_c$) view of the 3D volumes. Batch size of 1 was used during training, consisting of one sequence of three consecutive slices from the same patient, extracted from the same plane. To avoid overfitting, 3-fold cross validation was performed for $M_a$, $M_s$ and $M_c$, resulting in a total of 9 final models. A fixed maximum number of epochs was set to 150. Adam optimizer with a fixed learning rate equal to 0.0002 was used for training. 

\subsection{Model performance analysis}

Training and validation losses were monitored during training the network. During training, the mean DSC was calculated over the 3 validation sets, made of sequences selected using the criteria explained in section 2.3.1. The highest value of mean DSC was achieved in the first 40 epochs when trained on the selected sequences from the axial (0.72 (mean) ± 0.20 (std) DSC), coronal (0.68 (mean) ± 0.25 (std) DSC) and sagittal (0.72 (mean) ± 0.21 (std) DSC) planes of the validation set. Nevertheless, to obtain final predictions, we selected the last trained model after 150 epochs because the validation loss stabilized around a certain value and the fluctuations in both loss functions decreased.

\subsection{2D slices reconstruction results}

As a final step, 142 consecutive sequences were extracted from each image volume of each patient in the validation set and predictions were obtained. Slices from each volume were then reconstructed as explained in section 2.3.2. The performance of the method on each patient was assessed by comparing the mean DSC of the reconstructed slices for $M_a$, $M_s$ and $M_c$. We used different probability thresholds, when transforming images into binary masks, and we recalculated the evaluation metrics. Table \ref{direct2} shows the average results over all patients in the validation set for the DSC using a probability threshold of $th=0.9$. After the slice reconstruction, the mean DSC of the $M_s$ achieved the highest score and the mean DSC of the $M_a$ the lowest one.

\begin{center}
\begin{figure*}
\centering
\includegraphics[scale=0.60]{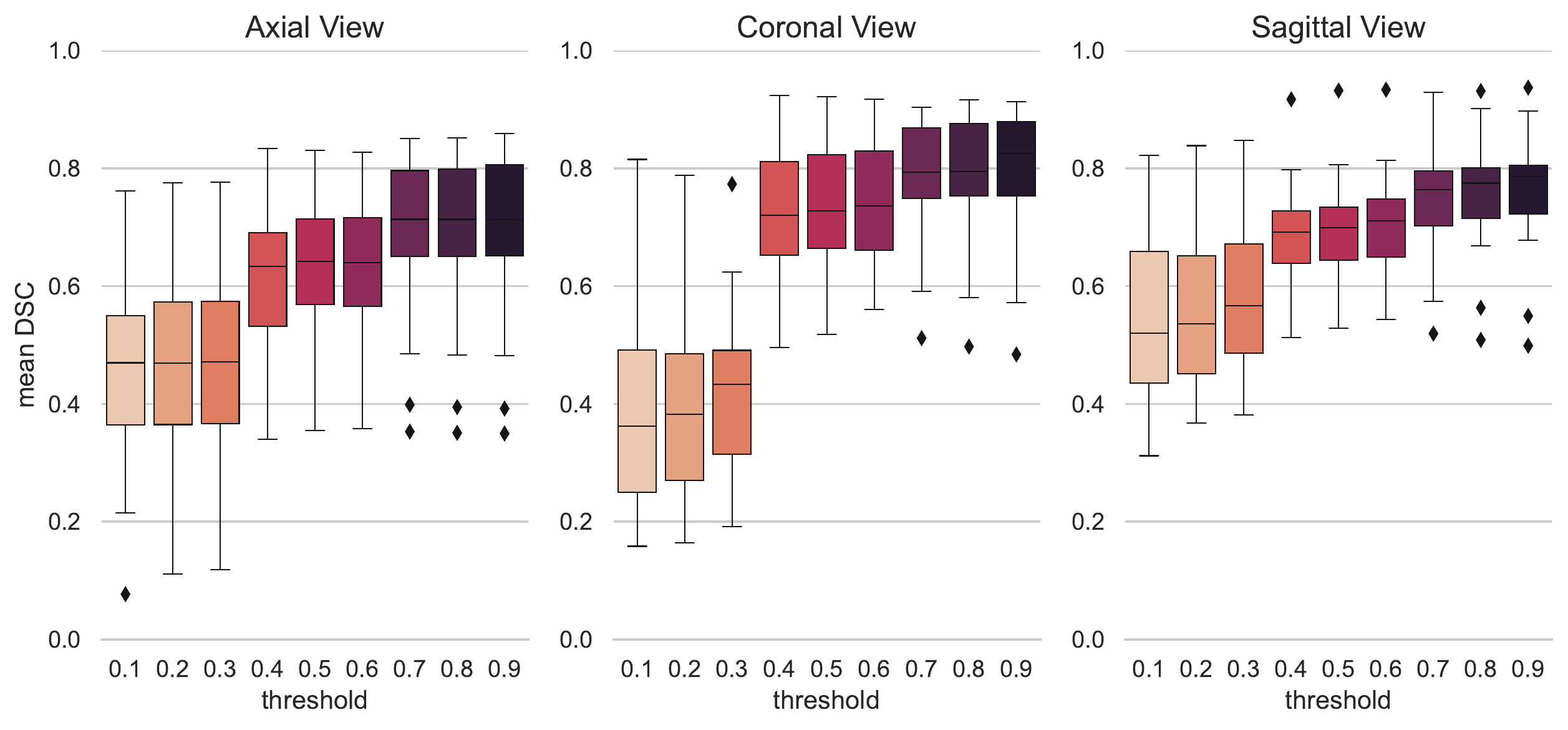}

\caption{Boxplots showing the mean DSC distributions of the patients in the test set at different probability thresholds. The plots report results obtained by ensembling the outputs of the models trained with sequences extracted from the axial, coronal and sagittal view. For the first and the third graph a similar ascending pattern can be observed from low to high thresholds of probability. For the coronal view, a bigger difference in results is shown in correspondence of $ th = 0.3 $ and $ th = 0.4 $.} \label{patients}

\end{figure*}
\end{center}

When testing, in each direction, slices were obtained by ensembling predictions of the three optimized models trained using 3-fold cross validation. 
In figure \ref{patients} the distribution of the mean DSC obtained from the test set is reported at different probability thresholds using boxplots. An ascending pattern can be observed, meaning that when the threshold of the predicted probability increases, the segmentation prediction overlaps better with the ground truth delineation of the tumor. For all views we observed a similar grouping pattern: boxplots behave similarly between threshold values of 0.1-0.3, 0.4-0.6 and 0.7-0.9. In the coronal view boxplots show a more distinctive difference in mean DSC distribution at $ th = 0.4 $. 
In table \ref{direct2} results on the test set are reported for the mean DSC. The values obtained on the validation and test set are consistent. The results from the models trained on sequences extracted from the coronal and sagittal view outperformed the ones from the axial in both the validation and test results. The mean DSC values in the test results confirmed the robustness of the ensemble technique. 
%\vspace*{-1.2cm}

 %The lowest value of median HD distance was obtained when training the network on slices extracted from the sagittal plane of the 3D volumes. 
%Only 2 out of 25 patients resulted in both zero precision and zero recall as result of $M_a$, $M_s$ and $M_c$. Hence, the network misses two patients belonging to the test set
%were consistent if looking at the DSC: high mean DSC corresponds to low median HD. 

\begin{table}
\centering
\caption{Mean DSC for the validation and test set. The validation outputs were obtained using 3-fold cross validation, while the test outputs were obtained by ensembling predictions of the three models optimized on each fold. For each patient, slices were reconstructed from predicted sequences of slices extracted from the axial, coronal and sagittal plane of the 3D FDG PET/CT images. A threshold value of 0.9 was chosen to make the outputs binary.}\label{direct2}
\centering
\vspace*{0.5cm}
\begin{tabular}{llcc}
\hline
\hline
  &   & \multicolumn{2}{c}{\small mean $\pm$ std} \\ 
Metric & model & \multicolumn{2}{c}{\footnotesize (min - max)} \\ 
\cline{3-4}
      &                     &  validation   &    test \\ [0.3ex] 
\hline
\hline
\rule{0pt}{4ex}    
 DSC  &   {\bfseries $M_a$} & 0.61$\pm$0.11 & 0.70$\pm$0.14  \\ %[-0.8ex] 
      &                     & \footnotesize	(0.22-0.82) & \footnotesize	(0.35-0.86) \\ %[0.3ex] 
      &   {\bfseries $M_c$} & 0.66$\pm$0.18 & {\bfseries 0.80}$\pm$0.11  \\ %[-0.8ex] 
      &                     & \footnotesize	(0.21-0.93) & \footnotesize	(0.48-0.91) \\ %[0.3ex] 
      &   {\bfseries $M_s$} & {\bfseries 0.68}$\pm$0.11 &  0.77$\pm$0.10 \\ %[-0.8ex] 
      &                     & \footnotesize	(0.34-0.87) & \footnotesize	(0.50-0.94) \\ %[0.3ex] 
\hline
\end{tabular}
\end{table}

%\hline\\
%\hline
%\hline
%Metric & model & \multicolumn{2}{c}{median} \\ [0.3ex] 
%\cline{3-4}
%         &       &      validation      &    test\\ [0.3ex]
%\hline
% 95\% HD  &   {\bfseries $M_a$} & 7.810 & {\bfseries 0}  \\ [0.3ex] 
%mm)     &   {\bfseries $M_c$} & 12.205 & {\bfseries 0}  \\ [0.3ex] 
%               &   {\bfseries $M_s$} & {\bfseries 0} & {\bfseries 0}  \\ [0.3ex]  

% mean train: 13.43832749	18.6745361	20.24046006
% / mean test: a 4.501 c 7.542 s 2.591
%\rule{0pt}{2.5ex}  {\bfseries not specified}  &  55  &  \\ [0.4ex] 

%\begin{center}
\begin{figure}%[!ht]
\centering
%\hspace*{-0.2cm}
\includegraphics[scale=0.55]{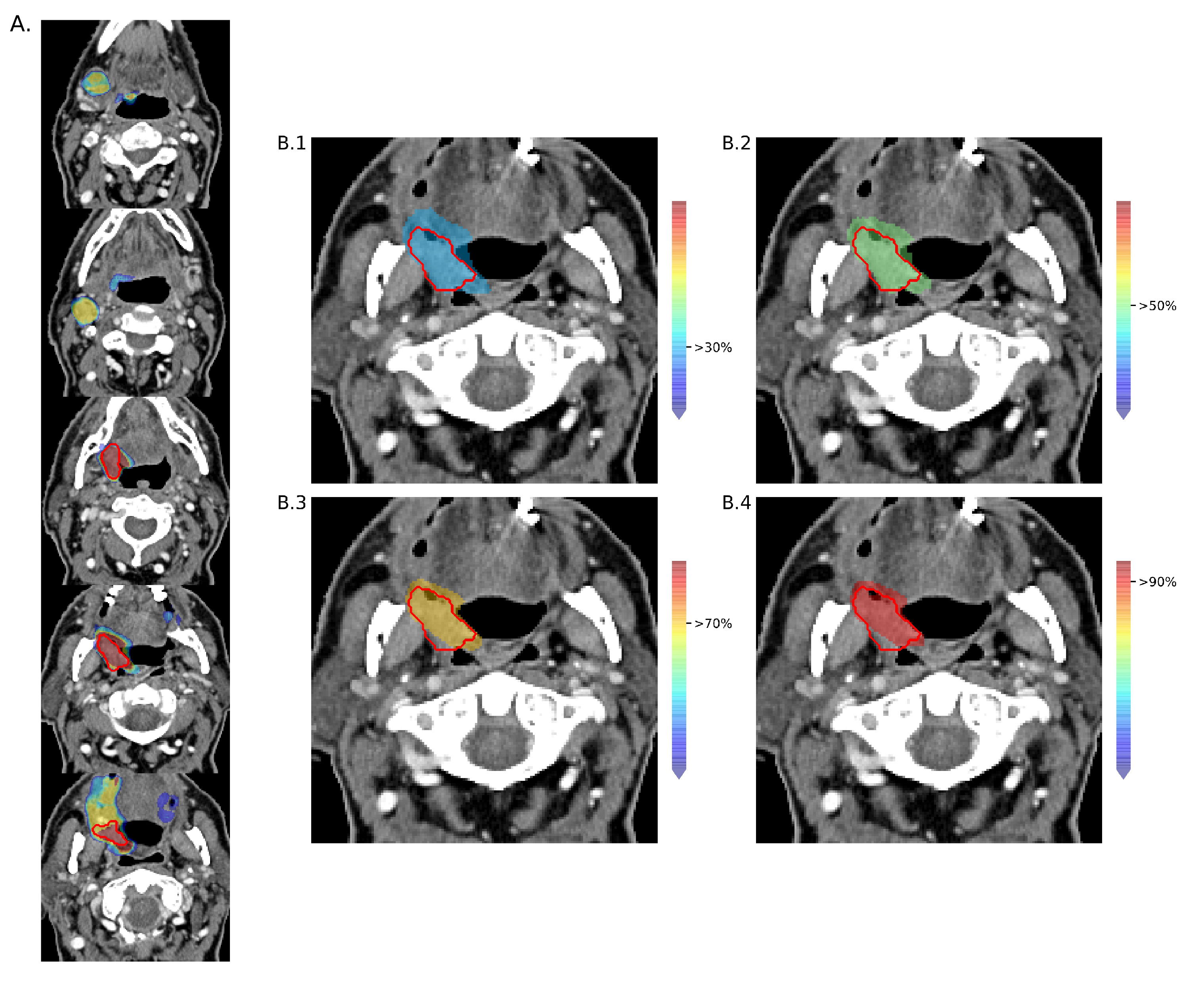}

\caption{Examples of the output of the proposed method. {\bfseries A.} Outcome probabilities displayed on the axial view of slices extracted from the CT 3D volume of a patient in the dataset. The red line represents the GTVp provided as ground truth, while the different colors show different areas of tumor probabilities. {\bfseries B.} Four examples of customized outputs based on different probability threshold settings, as shown by the color bars on the right side of the images. In {\bfseries B.1} only the areas containing pixels with tumor probabilities above 30\%  are displayed, above 50\% in {\bfseries B.2}, above 70\% in {\bfseries B.3} and above 90\% in {\bfseries B.4}.}\label{best_th}
%\vspace*{-2cm}

\end{figure}
%\end{center}

%One way ANOVA

In figures \ref{test_T} and \ref{test_N}, in the Appendix, boxplots showing the mean DSC distributions across probability thresholds for different T and N stages can be found. An increasing pattern in mean DSC can be observed for T1, T2, T3 and T4 stages in all three views. However, in the axial and coronal views, predictions of T1 stages considerably improve when choosing probability thresholds above 0.3. Predictions of T4 stages result always in higher mean DSC compared to predictions of lower T stages, except for high probability thresholds in the sagittal view. Low N stages show a similar increasing pattern of low T stages across thresholds. For the N3 stages mean DSC seems independent of the thresholds. 

\subsection{Qualitative results}

In figure \ref{best_th}, we show how the probability map predictions of our network looks on the CT image of a patient. On the left side, the total range of probability is shown on few slices extracted from the axial plane. On the right side, four cases of different probability threshold settings are displayed. The range of probability shown on the image for each case corresponds to the one above the selected threshold up to 100\%. Increasing the probability percentage shrinks the predicted area around the tumor. 

\begin{figure}%[!ht]
\centering
\hspace*{-0.4cm}
\includegraphics[scale=0.65]{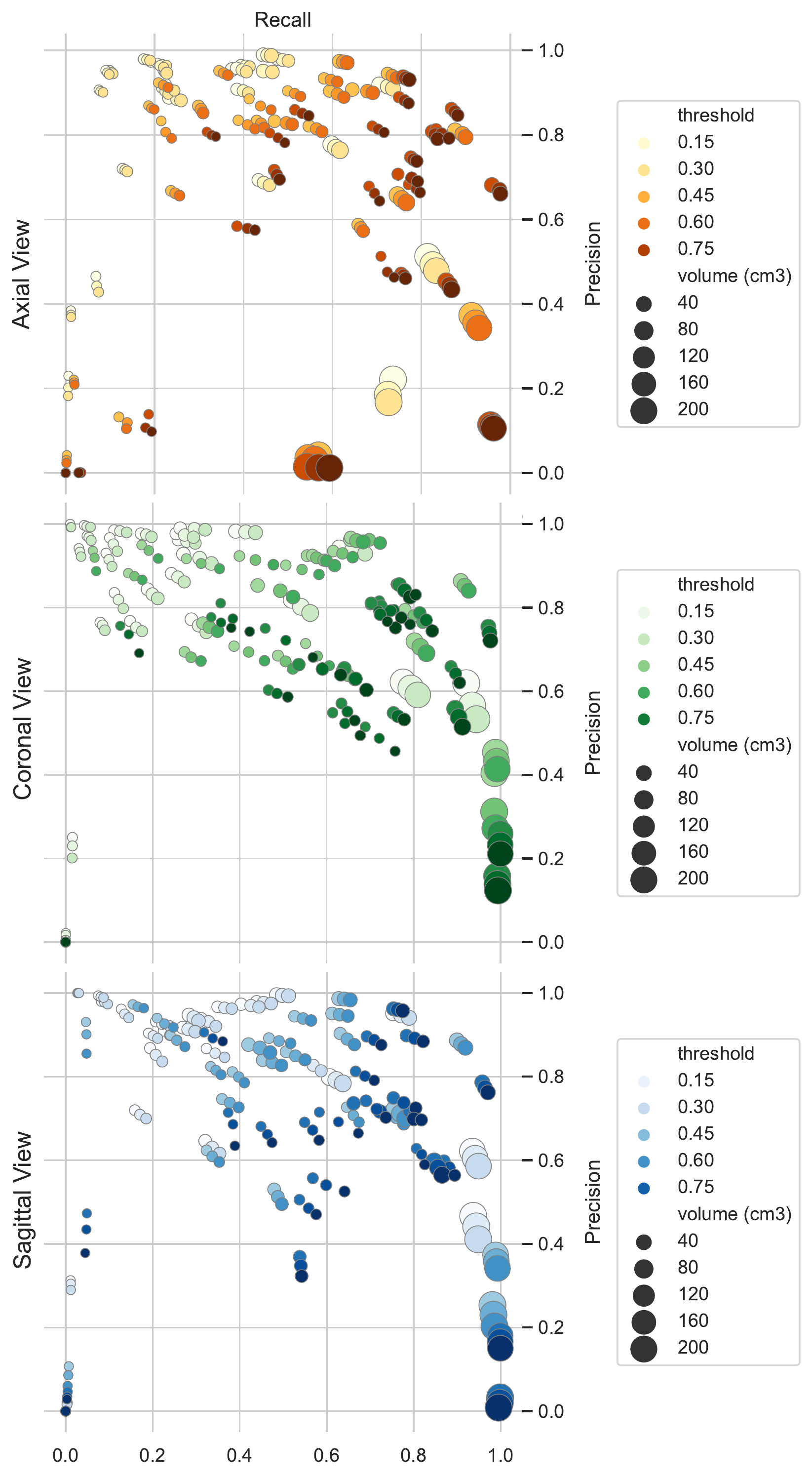}
\caption{Scatter plots of recall versus precision metric values for each patient in the test set. For each patient precision and recall values are calculated as the average of the metrics results obtained from the reconstructed slices. The size of each data point represents the tumor volume, while the color intensity represents the probability threshold chosen to make the image binary.}\label{prec_recall}
%\vspace*{-2cm}

\end{figure}

After a qualitative assessment of the results from the axial view, we acknowledged that the low performance of the models, in terms of mean DSC, corresponds to cases where big metastatic lymph nodes were present and tumor volume was larger. In figure \ref{prec_recall} we show the relationship between precision and recall per patient, at different probability thresholds, adding the tumor volume information from the reference contours by the radiation oncologists. For volumes reconstructed from sequences extracted from the coronal and sagittal view a similar pattern was observed. For patients with larger tumor volumes, precision is usually lower and recall is above 0.8. Having high recall on the image translates into having a low false negative rate. Precision in these cases is higher at low thresholds and decreases at higher thresholds of probability. The first scatter plot on figure \ref{prec_recall}, which represents results on the axial view, shows that for some of the patients with larger tumor volumes, values of recall are around 0.6. 

 %According to the patient files, for two patients in the dataset the tumor was visible but difficult to define because of nodal metastases. The predictions on the images show a central area containing a probability above 80\% and a surrounding area with lower probability values. For these slices the GTVp covers a large part of the 2D image.

\section{Discussion}

%https://www.arxiv-vanity.com/papers/2010.06163/
The experimental results show the potential of the proposed 2D deep learning adaptive segmentation method for oropharyngeal cancer in PET-CT images. During the second edition of the HECKTOR challenge \cite{Andrearczyk2022a}, Xie and Peng \cite{Xie2022} ranked first in the task of tumor segmentation using 3D PET-CT images, achieving a value of mean DSC of 0.7785 on a test set of 101 cases from two different centers.  The performance of our method in terms of mean DSC is not comparable with the 3D segmentation results achieved in the literature for the same task \cite{ Oreiller2022}, because in the calculation of the mean DSC we also included no tumor slices present in the selected bounding boxes. 
Values of precision and recall can be considered for a fairer comparison of the results. As discussed in \cite{Andrearczyk_first}, 2D methods result in higher precision and lower recall than 3D methods. High values of precision and recall in figure \ref{prec_recall} correspond to the top right corner in the box plots where tumors with smaller volumes appear. Our mean results in terms of precision range between 0.57 and 0.83 across thresholds, in all three different views, against 0.6241 of the 2D PET/CT baseline in \cite{Andrearczyk2020d}. In our study the mean recall ranges between 0.5 and 0.67 at probability thresholds above 0.5, which is slightly lower than the 2D PET/CT baseline in \cite{Andrearczyk2020d}.

In this paper we proposed a 2D deep learning adaptive segmentation method that aims to assist radiation oncologists in segmenting primary tumors. The framework mimics experts in their task of slice-by-slice tumor contouring on PET and CT images of oropharyngeal cancer patients. The added value of the method is providing probability maps which could be used in form of clinical decision support as a starting point in the contouring process by the radiation oncologist. Providing the radiation oncologist with contours of the GTV primary tumor for different probabilities, makes it possible to choose the most appropriate contour. This contour could than be optimized based on clinical information and additional imaging. It is expected that by offering the radiation oncologist a GTV contour to start with, the contouring process can become faster and the inter observer variability could be reduced \cite{Vaassen2020}.

Despite several studies showing promising results in recognizing head and neck tumors on multi-modal image data using deep learning \cite{Oreiller2022}, the limitations involved are still an obstacle for their clinical implementation. 
In this paper, we suggest a solution that tackles the problem of high variability in ground truth images used for tumor segmentation, and that may enhance the utility of such a tool in clinical practice. The proposed method, in fact, displays on different areas on the image a tumor probability derived by the tumor segmentation learning process and supports the end user towards a quicker and more attentive decision making. In tasks where the gold standard is not represented by one and only one ground truth, like the one of tumor segmentation, a large number of different annotations would be required for training a DL network. In several deep learning studies, one adopted solution, is to collect consistent tumor delineations obtained after consultations of different radiation oncologists for the whole dataset. The dataset used, in these cases, is consistent, and the deep learning framework learns according to what has been decided to be "the truth". In our paper, data has been collected between 2014 and 2017. In this time period, different radiation oncologists have been working in our center and additional image modalities became available for a more precise final tumor contour (e.g. MRI). Despite the delineation guidelines being the same, the variation among tumor delineations is large. Supervised learning methods reproduce human biases and errors, thus the problem of inter-observer variability in target volume delineation is a perpetuated error. This can be one of the reasons why our results are sometimes lower than in other studies and hard to be compared with. 

However, relabeling a dataset is not practical and it is time consuming. New data will always be collected and used to update existing models and similar problems will again be encountered. In our method, in fact, we capture the variation in contouring in the training set and we display the variation in contours between the different probability threshold values. To widen the output probability range and to obtain more robust predictions, ensembling learning was performed. The ensambling technique provided more stable results, as shown by the lower standard deviation of the test set results in table \ref{direct2} compared to the validation set results. To evaluate the method presented in this paper, we calculated mean DSC at the different probability thresholds to establish whether the outcome predictions were overlapping the tumor area or not. Increasing the probability threshold, predictions showed to become more similar to what the oncologist defined as tumor delineation. Since the segmentation task was performed on 2D images, the evaluation results for each patient in each spatial direction were obtained as the average dice score coefficient calculated between each 2D slice and the corresponding ground truth. The same method was used also for precision and recall. When setting a probability threshold of 0.9, only two out of 25 tumors have a corresponding precision and recall below 0.03. In both cases the tumor dimension was small and really difficult to identify on PET, as also mentioned in the patient file.
%The framework was trained on 3 different folds of the same dataset during cross validation. All models were saved and used when testing on the test set of 25 patients. The final outcome results were obtained as the averaged probability maps of the predictions.

%add the registration problems
In our work, we build on the fact that tumors are identified on a pixel level on multi-modal images. The combination of PET and CT images gives complementary information that makes the network assign a certain tumor probability to a pixel. The channel attention blocks of the first part of the framework (in figure \ref{workflow}) are intended to guarantee cross-modal learning. On one hand, different imaging modalities have a positive impact on contouring consistency, taking advantage of all the imaging information available for segmentation \cite{Bird2015}. On the other hand, contour variations can result from differences in tumor visualization in various imaging modalities and not accurate multi-modal image co-registrations \cite{Sadeghi2021}. PET imaging, in fact, does not allow a high delineation accuracy but is fundamental in identifying head and neck cancer. The implemented deep learning framework heavily relies on FDG uptake in PET images. Since the brain area is characterized by a large amount of pixels with high SUV values, these areas also appear in the outcome prediction, but with low probabilities. This explains the large difference in mean DSC between probability thresholds below and above 0.3 in the coronal view. When trained on sequences from the axial plane, the network appears to rely on PET intensity and on tumor location. In fact, in presence of primary tumors with larger volume that expand to metastatic lymph nodes (higher N stages), the mean DSC values result lower. For the purpose of this study, metrics need to be calculated at different thresholds of probability to have an overall idea of the method performance. A balance between precision and recall is what any model aims for, but in this study a lower precision and an higher recall can indicate the volume at risk for tumor extension, which should be further explored by the radiation oncologist.

%clinical - utility of such a tool in clinical practice:
%The deep learning framework does not aim to substitute radiation oncologists, but to be of help in decision making. Image segmentation tasks 

%reasons why mean DSC and HD are lower than in other studies

\section{Conclusions}
In conclusion, this study proposes a novel deep learning aided tool for oropharyngeal tumor segmentation on FDG PET/CT images using probability maps. The framework is designed to capture cross-modality information and inter-slice context using spatial attention and bi-directional  long  short term memory mechanisms. The main object is to display tumor probabilities on 2D slices extracted from the axial, sagittal and coronal view of the 3D image volumes, to assist radiation oncologists in decision making during radiotherapy planning. The available wide range of probabilities is of interest to minimise the risk of geographical misses and to make available the uncertainty derived from the tumor segmentation task, leading cause of inter-observer variability. Experimental results show the potential of the proposed method in identifying the primary tumor.

\section*{Availability of data and materials}
The data used for this research comprise confidential patient health information, which is protected and may not be released unless approved by the Committee of Ethics of the UMCG.

\section*{Declaration of competing interest}
The authors declare no conflict of interest.

\section*{Acknowledgment}

This research was supported by the Hanarth Fonds. We would like to acknowledge the contribution in data collection of the PhD students working on Head and Neck cancer research in the Department of Radiation Oncology, at the University Medical Hospital of Groningen, Netherlands. 
We would like to thank the Center for Information Technology of the University of Groningen for their support and for providing access to the Peregrine high performance computing cluster.

\bibliographystyle{elsarticle-harv} 
\bibliography{Mysecondpaper.bib}


%\begin{thebibliography}{2} %00

\newpage
\section*{Appendix}

%\begin{center}
%\begin{figure}
%\centering
%\includegraphics[scale=0.58]{images/meanprec_test.pdf}
% \includegraphics[width=\textwidth]{housd.pdf}

%\caption{Mean Precision of patients in test set.} \label{test_meanprec}
%\vspace*{-1.5cm}

%\end{figure}
%\end{center}    

%\begin{center}
%\begin{figure}
%\centering
%\includegraphics[scale=0.58]{images/meanrec_test.pdf}
% \includegraphics[width=\textwidth]{housd.pdf}

%\caption{Mean Recall of patients in test set.} \label{test_meanreca}
%\vspace*{-1.5cm}

%\end{figure}
%\end{center}

\begin{figure}[H]
\centering
\includegraphics[scale=0.5]{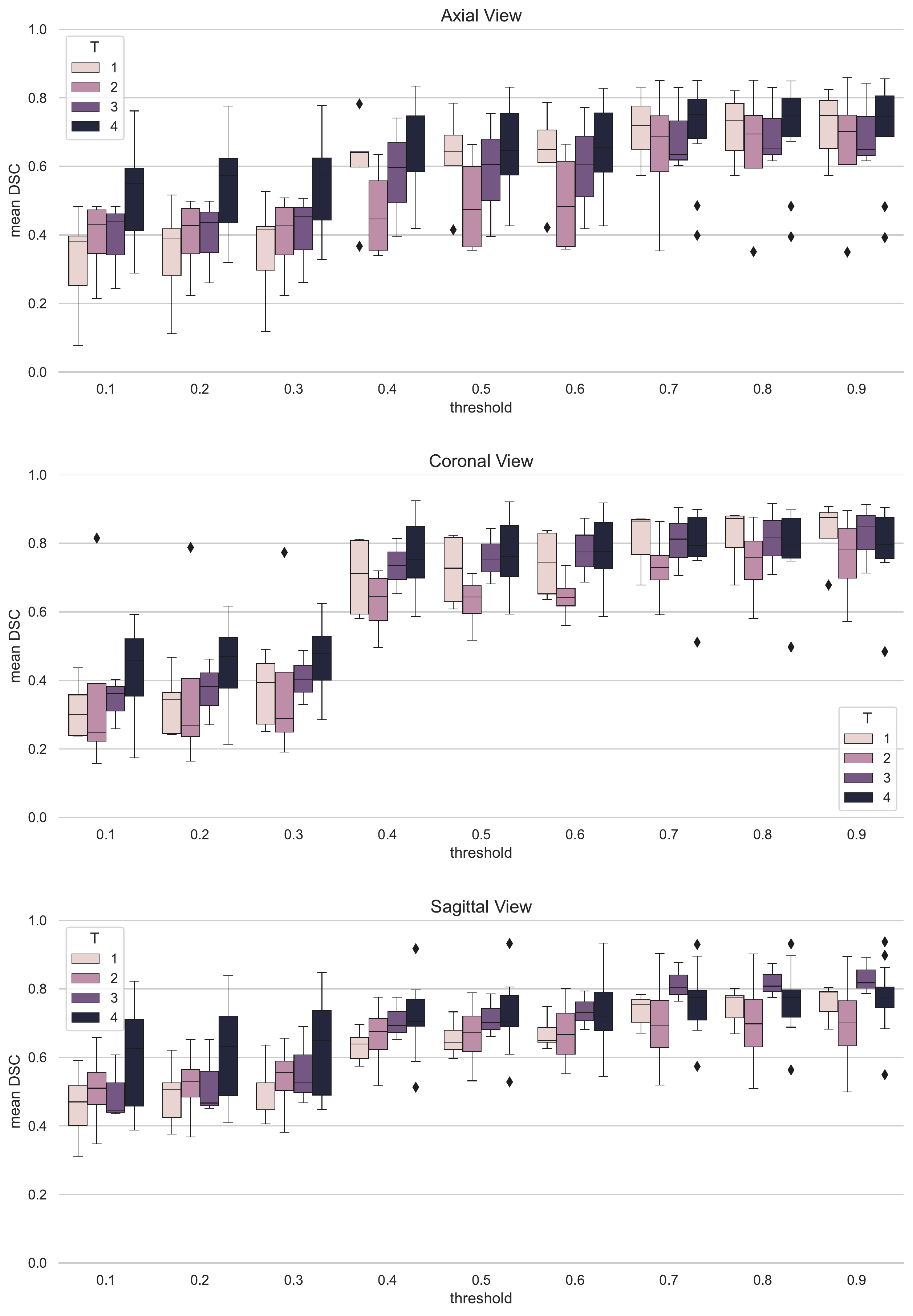}

\caption{Mean DSC of patients in test set grouped by T stages at different probability thresholds.} \label{test_T}
%\vspace*{-1.5cm}

\end{figure}

\vspace*{-1.5cm}

\begin{center}
\begin{figure}
\centering
\includegraphics[scale=0.5]{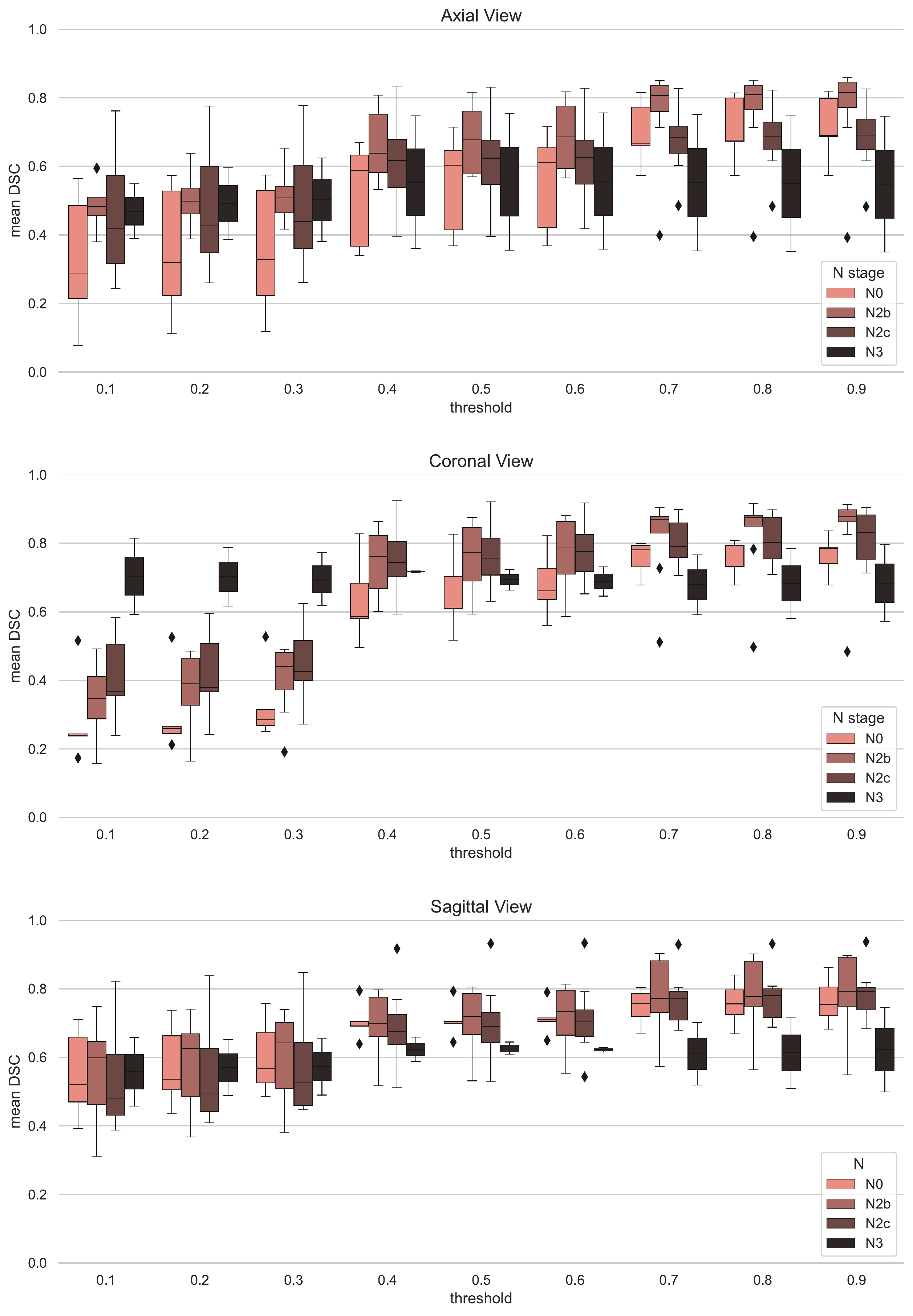}

\caption{Mean DSC of patients in test set grouped by N stages at different probability thresholds.} \label{test_N}
%\vspace*{-1.5cm}

\end{figure}
\end{center}

\end{document}